\documentclass[aps,prl,groupedaddress,showpacs,notitlepage,twocolumn]{revtex4-1}
\usepackage{hyperref}
\usepackage{amsmath}
\usepackage{graphicx}
\usepackage{color}

\usepackage[T1]{fontenc}
\usepackage[latin1]{inputenc}

\newcommand{\be}{\begin{equation}}
\newcommand{\ee}{\end{equation}}

\begin{document}

\title{Edge Solitons in Nonlinear Photonic Topological Insulators}

\author{Daniel Leykam$^1$}

\author{Y.~D.~Chong$^{1,2}$}

\affiliation{$^1$Division of Physics and Applied Physics, School of Physical and Mathematical Sciences,\\
Nanyang Technological University, Singapore 637371, Singapore\\
$^2$Centre for Disruptive Photonic Technologies, Nanyang Technological University, Singapore 637371, Singapore}

\date{\today}

\begin{abstract}
  We show theoretically that a photonic topological insulator can support edge solitons that are strongly self-localized and propagate unidirectionally along the lattice edge.  The photonic topological insulator consists of a Floquet lattice of coupled helical waveguides, in a medium with local Kerr nonlinearity.  The soliton behavior is strongly affected by the topological phase of the linear lattice.  The topologically nontrivial phase gives a continuous family of solitons, while the topologically trivial phase gives an embedded soliton that occurs at a single power, and arises from a self-induced local nonlinear shift in the inter-site coupling.  The solitons can be used for nonlinear switching and logical operations, functionalities that have not yet been explored in topological photonics.  We demonstrate using solitons to perform selective filtering via propagation through a narrow channel, and using soliton collisions for optical switching.  
\end{abstract}



\maketitle

Topologically nontrivial photonic bands, analogous to electronic topological insulators, have now been realized and studied in a variety of photonic structures~\cite{topological_photonics_review,Raghu1,Raghu2,wang2009,khanikaev2013,rechtsman2013,hafezi2013,metamaterial,plasmons,mukherjee2016}.  These ``photonic topological insulators'' (PTIs) feature edge states that are topologically protected against certain classes of disorder, and have interesting potential applications as robust waveguides and delay lines.  Thus far, PTIs have mostly been studied in the linear limit, in which existing concepts of band topology can be directly applied to the electromagnetic wave equations, such as by mapping the propagation equations for a linear photonic lattice to the linear Schr\"odinger equation~\cite{rechtsman2013}.  Even recent studies of PTIs arising in optically nonlinear systems, such as exciton-polaritons in quantum wells, have focused on topological edge states that are linear perturbations around a steady-state nonlinear background~\cite{bardyn2016,chi2,galilo_2015}.  There have been only a handful of investigations into the \textit{non-perturbative} nonlinear dynamics that could arise in PTIs~\cite{TI_solitons,ablowitz2014,nonlinear_SSH,nonlinear_QW,PTI_MI}.  Notably, Lumer \textit{et al.} discovered a localized stationary soliton lying in the bulk of a two-dimensional (2D) PTI, which can be interpreted as a point region of a different topological phase that is ``self-induced'' by topological edge states circulating around it~\cite{TI_solitons}.  Ablowitz \textit{et al.}~have found evidence for moving edge solitons in weakly nonlinear 2D PTIs, though this was done by taking broad envelope superpositions of existing topological edge states, and reducing the system to a 1D nonlinear Schr\"odinger equation~\cite{ablowitz2014}.  In 1D lattices, nonlinear dynamics of boundary states and self-induced topological transitions have also been studied~\cite{nonlinear_SSH,nonlinear_QW}.

This paper describes a class of moving lattice edge solitons that arise in experimentally feasible 2D PTIs with Kerr nonlinearity.  Unlike in Ref.~\cite{ablowitz2014}, the solitons are derived \textit{ab initio}, without using broad envelope approximations, in a realistic photonic lattice; furthermore, they can arise whether the underlying lattice is topologically trivial or nontrivial in the linear limit.  The underlying topological phase strongly affects the soliton properties.  In the topologically trivial phase, where the linear lattice lacks topological edge states, the solitons are topologically self-induced, similar to the stationary solitons found in Ref.~\onlinecite{TI_solitons} except that these can move unidirectionally along the edge.  They appear to be ``embedded'' lattice solitons, meaning that they co-exist with extended linear modes without being stabilized by a gap~\cite{embedded_solitons_1,embedded_solitons_2,embedded_solitons_3}; the soliton solution occurs at a single power, at which the radiative loss via coupling to small-amplitude linear waves happens to vanish.  We note that the only embedded lattice solitons experimentally observed so far have been stationary~\cite{embedded_solitons_expt}; well-localized moving lattice solitons are predicted to exist based on discrete models \cite{embedded_solitons_2,Pelinovsky2005,Malomed2006}, but have been challenging to realize experimentally.  On the other hand, when the underlying lattice is topologically nontrivial, moving edge solitons occur over a continuous range of powers, and are stabilized by the dispersion features of the linear topological edge modes.

The existence of strongly-localized moving solitons opens up a range of interesting possiblities for performing signal processing in PTI lattices~\cite{signal_processing}, beyond what could be accomplished using linear topological edge states or stationary solitons \cite{TI_solitons}.  We present two representative examples: (i) self-focusing of edge modes, allowing them to be ``squeezed'' through narrow channels without back-reflection, and (ii) collisions between edge and bulk solitons, which can be used for all-optical signal switching.

\begin{figure}
\includegraphics[width=0.9\columnwidth]{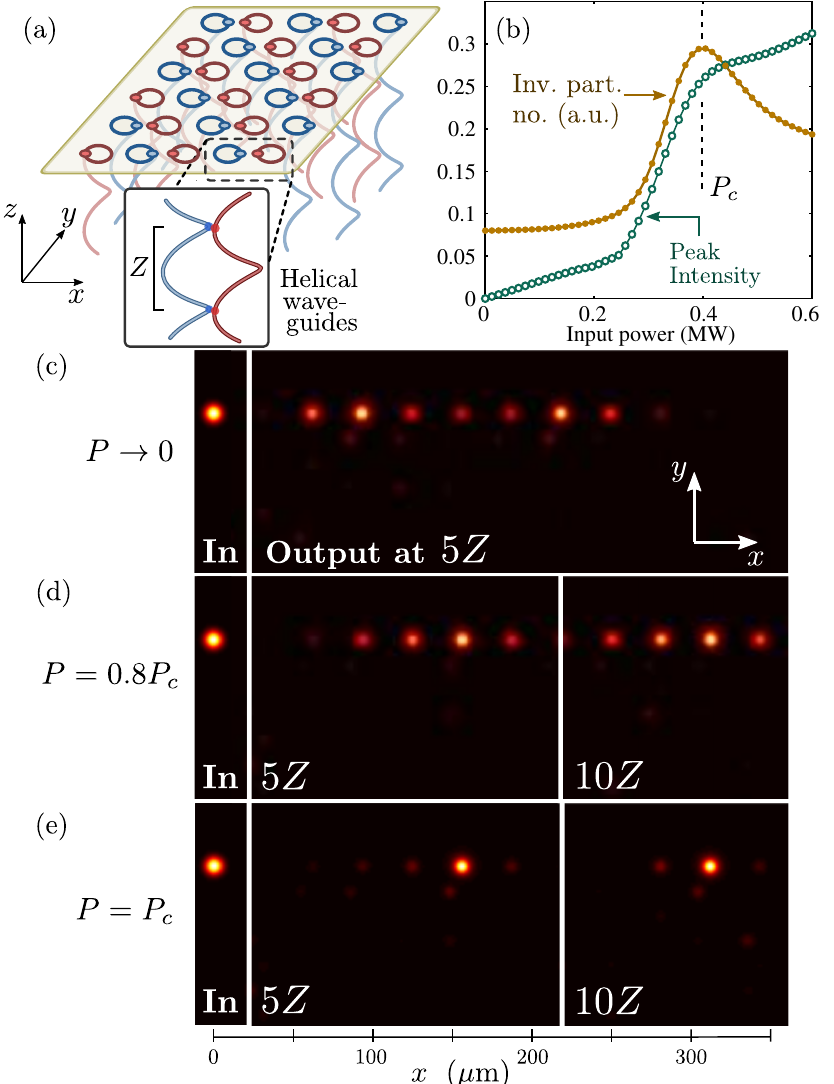}
\caption{(Color online) (a) Schematic of a photonic lattice of helical waveguides forming a square lattice in the $x$-$y$ plane.  On each sublattice (red and blue), the helix phase is staggered by $\pi$.  (b)--(e) Beam propagation simulation results showing solitonic self-focusing in a topologically nontrivial lattice.  The lattice consists of a strip $4$ lattice constants wide, with single-site input on the edge at $z = 0$.  (b) Peak intensity $I_{\mathrm{max}}$ and inverse participation number $\mathcal{P}^{-1}$ at $z = 5Z$ (i.e., after 5 helix cycles), versus input power $P$.  Corresponding in-plane intensity profiles at various propagation distances, for (c) $P\rightarrow 0$, (d) $P = 0.8P_c$, and (e) $P = P_c$, where $P_c$ is the power at which the soliton is most localized (i.e., $\mathcal{P}^{-1}$ is largest). }
\label{fig:nontrivial_propagation}
\end{figure}

The photonic lattice, shown schematically in Fig.~\ref{fig:nontrivial_propagation}(a), consists of a 2D square lattice of helical waveguides, ``staggered'' so that neighboring waveguides (in different sublattices) have helix phase shifts of $\pi$ relative to each other; hence, each waveguide approaches its four neighbours sequentially during each helix cycle~\cite{leykam2016}.  The waveguides are otherwise identical.  In the linear optics regime, we have previously shown that the photonic bandstructure can be either topologically trivial (a conventional insulator) or nontrivial (an anomalous Floquet PTI), depending on the inter-waveguide couplings~\cite{leykam2016}.  Now, we include a local Kerr nonlinearity in the optical medium.  In the paraxial approximation, the propagation of a monochromatic beam envelope $\psi(x,y,z)$ obeys
\be 
i \partial_z \psi = -\frac{1}{2k_0} \nabla^2_{\perp} \psi - \frac{k_0}{n_0} \Big(n_L (x,y,z) + n_2 |\psi|^2 \Big) \psi. \label{eq:nlse}
\ee
This is a nonlinear Schr\"odinger equation with the axial coordinate $z$ playing the role of time; $k_0 = 2 \pi n_0 / \lambda$ is the wavenumber, $\nabla^2_{\perp} \equiv \partial_x^2 + \partial_y^2$ is the transverse Laplacian, $n_0$ is the background refractive index, and $n_L$ and $n_2|\psi|^2$ are the linear and nonlinear local refractive index deviations from $n_0$.  The function $n_L(x,y,z)$ corresponds to the helix lattice described above, with mean waveguide spacing $a$, helix radius $R_0$, and modulation period $Z$.  We adopt parameter values consistent with fused silica glass at wavelength $\lambda = 1550\,$nm: $n_0 = 1.45$, $n_L \sim 2.7 \times 10^{-3}$ in the waveguides (and $n_L = 0$ outside), and $n_2 = 3 \times 10^{-20}$m$^2$/W~\cite{szameit_soliton} (self-focusing nonlinearity; a defocusing nonlinearity can give rise to similar effects~\cite{supplementary}).  The waveguides have circular cross sections with radius $4\,\mu$m.  The paraxial beam intensity $|\psi|^2$ is normalized by characteristic intensity $I_0 = 10^{16}$W/m$^2$, for which the nonlinear index shift $n_2 I_0$ is comparable to $n_L$. For the modal area of one waveguide $w_0^2 \sim (10\mu$m)$^2$, this requires peak powers $\sim 1$MW, accessible with pulsed lasers~\cite{szameit_soliton}.

In the linear regime ($n_2 \rightarrow 0$), the Floquet eigenmodes of the system can be obtained by solving
\be
\hat{U}(Z) \psi = \exp \left[ -i {\textstyle \int_0^Z} \hat{H} (z) dz \right] \psi = e^{-i \beta Z} \psi,
\ee
where $\hat{H}$ is the Hamiltonian in Eq.~\eqref{eq:nlse}, and $\beta$ is the Floquet quasienergy defined modulo $2\pi/Z$~\cite{kitagawa2010,rudner2013,pasek2014,rechtsman2013,leykam2016}.  The topology of the Floquet bandstructure is governed by an effective coupling angle $\theta_0 \in [0,\pi]$ controlling the band gap size, with the system topologically nontrivial when $0.25\pi < \theta_0 < 0.75\pi$ \cite{supplementary,leykam2016}.

We use beam propagation simulations of Eq.~(\ref{eq:nlse}) to study nonlinear waves produced by injecting light at the edge of a semi-infinite strip.  First, consider a latttice that is topologically nontrivial in the linear limit, with coupling angle $\theta_0 = 0.6\pi$ (from helix parameters $a = 22\mu$m, $R_0=4\mu$m, and $Z=1$cm).  When the input power $P$ is very low, the initial single-site excitation couples to linear topological edge states. As shown in Fig.~\ref{fig:nontrivial_propagation}(c), this produces a wavepacket propagating in one direction along the edge (since the edge states are unidirectional), while undergoing broadening (since the dispersion of the edge states is not perfectly linear~\cite{leykam2016}).  Upon increasing $P$, the Kerr nonlinearity induces self-focusing, and we observe solitonic propagation along the edge of the lattice, as shown in Fig.~\ref{fig:nontrivial_propagation}(d)--(e).  We momentarily put aside the issue of soliton stability, which will be discussed below.  To quantify the self-focusing, Fig.~\ref{fig:nontrivial_propagation}(b) plots the peak intensity $I_{\mathrm{max}} = \mathrm{max}(|\psi|^2$) and inverse participation number $\mathcal{P}^{-1} = \int |\psi|^4 \; dx dy / (\int |\psi|^2 dx dy)^2$, after propagation through 5 helix cycles.  At a critical power $P_c$, the soliton becomes localized to almost a single site, as shown in Fig.~\ref{fig:nontrivial_propagation}(e).  This self-focusing effect does not appear to be describable using a purely on-site nonlinearity of the sort used in Refs.~\onlinecite{TI_solitons,ablowitz2014}, but it can be modeled by a nonlinear shift in the effective coupling angle~\cite{nonlinear_coupler,nonlinear_coupler_2}:
\begin{equation}
  \theta_{\mathrm{eff}} \approx \theta_0 - \theta_{NL} |\psi|^2.
  \label{nonlinearity}
\end{equation}
In previously-studied discrete nonlinear lattice models, similar coupling nonlinearities were shown to produce ``compact'' solitons that are perfectly localized to a few sites~\cite{compacton,discrete_nonlinear_coupling}.  In our photonic lattice, the shift is due to the self-focusing Kerr nonlinearity increasing the waveguide depth, thus reducing the evanescent inter-waveguide coupling.  When $P = P_c$, the coupling angle reaches $\theta_{\mathrm{eff}} = 0.5\pi$, which corresponds to perfect coupling into neighbouring waveguides at each quarter-cycle of the helix.  This is the ``middle'' of the topological insulator phase, where the edge states are maximally localized.

\begin{figure}
\includegraphics[width=0.9\columnwidth]{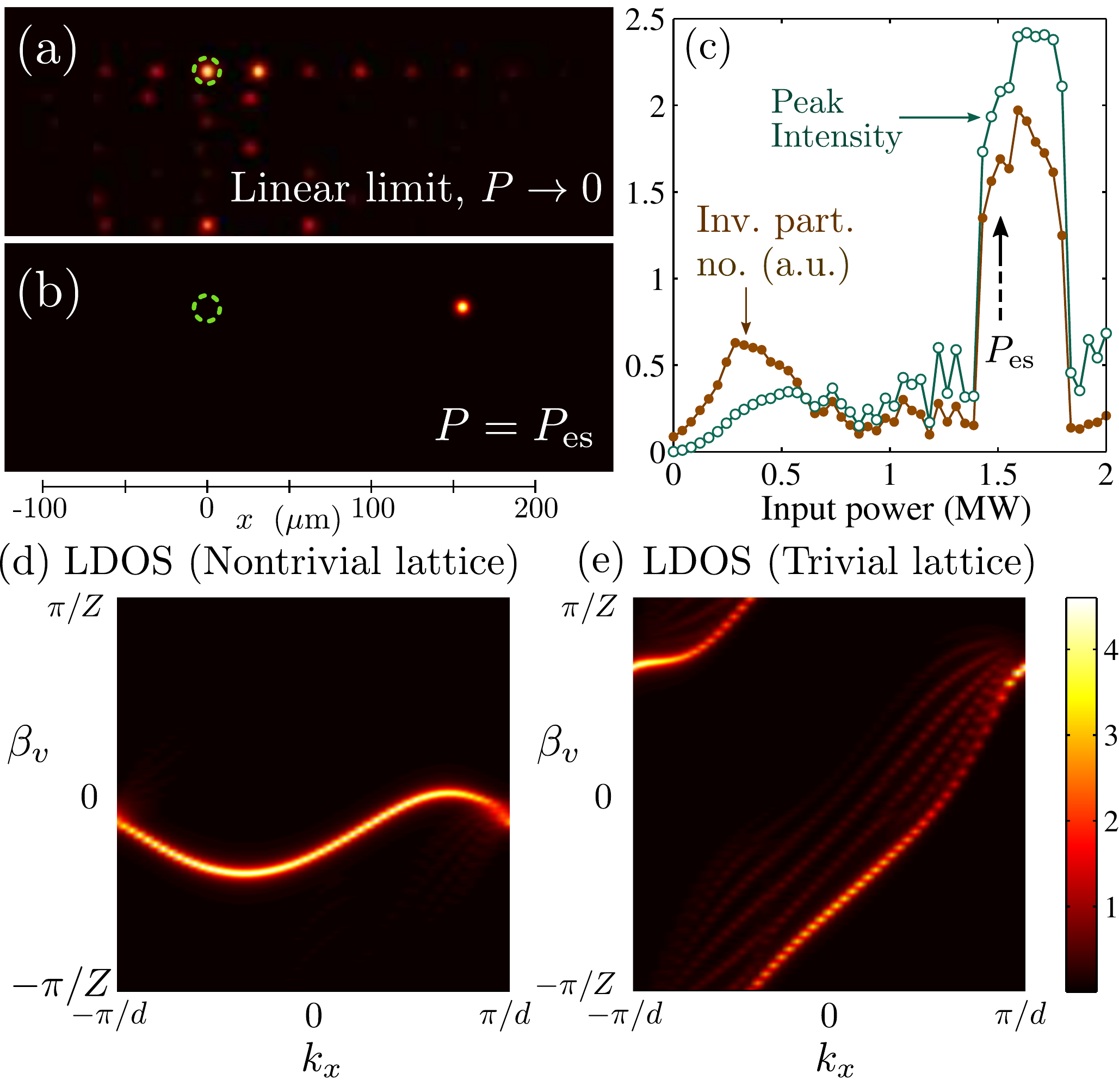}
\caption{(Color online) Self-induced nonlinear edge modes in a topologically trivial lattice. (a)--(b) Intensity profiles after propagation through $z = 5Z$, starting from a single-site excitation (dotted circle), for (a) the zero-power (linear) limit $P \rightarrow 0$, and (b) input power $P_{\mathrm{es}}$ corresponding to the embedded soliton.  (c)  Peak intensity $I_{\mathrm{max}}$ and inverse participation ratio $\mathcal{P}^{-1}$ at $z = 5Z$, versus input power $P$.  (d)--(e) Local density of states along the edge, for the nontrivial and trivial lattices in the linear limit, plotted using the moving-frame quasienergies defined by $\beta_v (k_x) \equiv \beta (k_x) - v k_x$.}
\label{fig:trivial_propagation}
\end{figure}

We now turn to a lattice that is topologically trivial in the linear limit.  As shown in Fig.~\ref{fig:trivial_propagation}(a)--(b), injecting light at the edge of the nonlinear lattice at a certain input power can produce a strongly-localized moving soliton.  Since the linear lattice in this case lacks topological edge states, the soliton in Fig.~\ref{fig:trivial_propagation}(b) is ``topologically self-induced'': the field intensity near the soliton drives the local region of the lattice from the topologically trivial phase to the nontrivial phase via Eq.~(\ref{nonlinearity}).  Here, the linear lattice has coupling angle $\theta_0 = 0.9\pi$ (from helix parameters $a=25\mu$m, $R_0 = 6\mu$m, and $Z=2$cm), and the nonlinearity drives the local coupling angle below the topological transition located at $\theta = 0.75\pi$.  Unlike the topologically self-induced stationary solitons observed in Ref.~\onlinecite{TI_solitons} (in a non-staggered helical lattice), these solitons are mobile, and move unidirectionally along the edge.

In contrast to the solitons found in the nontrivial lattice, the soliton in Fig.~\ref{fig:trivial_propagation}(b) is only observed near a specific input power $P_{\mathrm{es}}$.  Fig.~\ref{fig:trivial_propagation}(c) plots the peak intensity and inverse participation number versus input power $P$.  Both quantities are strongly suppressed except near $P_{\mathrm{es}} \approx 1.5$MW.  For other values of $P$, the input light diffracts strongly into the bulk, and the wavepacket decays with $z$ until the nonlinearity becomes negligible~\cite{surface_soliton}.

The stability of a soliton depends on whether there exist small-amplitude linear waves that the soliton can decay into.  For stable propagation at velocity $v$, the soliton must avoid resonance with linear waves in its moving frame, whose quasienergies are $\beta_v (k_x) \equiv \beta (k_x) - v k_x$~\cite{embedded_solitons_2}.  In Fig.~\ref{fig:trivial_propagation}(d)--(e), we plot the moving-frame local density of states (LDOS) on the lattice edge.  For the previous topologically nontrivial lattice (corresponding to results shown in Fig.~\ref{fig:nontrivial_propagation}), we see from Fig.~\ref{fig:trivial_propagation}(d) that the LDOS is dominated by the linear topological edge states (the bulk states' contribution is about 1\% of the bulk contribution in the trivial lattice).  Since the edge states are unidirectional and have nearly linear dispersion, they only occupy a narrow range of $\beta_v$.  This allows soliton families to avoid resonating with linear modes~\cite{ablowitz2014}. In the topologically trivial lattice, however, there is significant LDOS for all $\beta_v$, because the contributing bulk modes are not unidirectional. Thus, a traveling edge wave is typically destabilized by decaying into small-amplitude bulk modes.  As an exception, at a specific power $P_{\mathrm{es}}$ coupling to the bulk modes could vanish, leading to the formation of an ``embedded soliton'' \cite{embedded_solitons_1,embedded_solitons_2,embedded_solitons_3}. This explains the semi-stable behavior shown in Figs.~\ref{fig:trivial_propagation}(b)--(c):  embedded solitons typically have a finite basin of attraction, so for $P \gtrsim P_{\mathrm{es}}$ the excess energy is radiated away, but if $P$ is too large, the wavepacket fails to relax to the soliton.  

To further verify that these are soliton solutions, we formulated a discrete model for the photonic lattice and searched for nonlinear modes using the Floquet self-consistency method developed in Ref.~\onlinecite{TI_solitons}.  In the nontrivial lattice, the procedure converged numerically to solutions corresponding to moving solitons on the edge, as well as stationary gap solitons in the bulk.  In the trivial lattice, we obtained convergence to an embedded moving soliton localized to a single site, and verified that its power can be described by $P_{\mathrm{es}} \approx (\theta_0 - \pi/2)/\theta_{NL}$.  Details are given in the Supplemental Material \cite{supplementary}.

\begin{figure}
\includegraphics[width=\columnwidth]{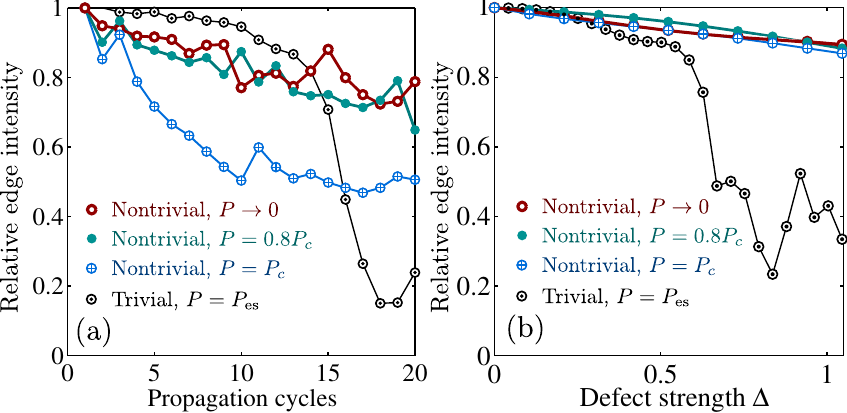}
\caption{(Color online) Stability of nonlinear edge modes in trivial and nontrivial photonic lattices.  (a) Relative edge intensity (the ratio of the edge intensity to the initial edge intensity at $z=0$) versus propagation distance $z$.  (b) Relative edge intensity at $z = 5Z$, after scattering off a defect with normalized strength $\Delta \equiv \delta n k_0 Z / (2\pi n_0)$, where $\delta n$ is the detuning of the defect waveguide's refractive index. }
\label{fig:stability}
\end{figure}

Another practical constraint on soliton stability comes from bending losses, i.e.~radiation into non-lattice modes.  For the chosen helical waveguide parameters, bending losses for the linear topological edge states are very low, on the order of 1\% per helix cycle~\cite{leykam2016}.  Fig.~\ref{fig:stability}(a) shows the power retained on the edge waveguides with increasing $z$.  In the nontrivial lattice, after an initial power-dependent transient oscillation (caused by the choice of single-waveguide excitation), the solitons have decay rates comparable to the linear topological edge states.  In the trivial lattice, however, the soliton abruptly destabilizes after $\sim 15$ cycles, due to bending losses reducing the power below $P_{\mathrm{es}}$ and triggering a breakup.

We now ask how robust the solitons are against defects and lattice shape deformations.  Topological edge states of linear PTIs are known to be highly robust, limited mainly by finite-size effects (i.e., scattering to the opposite edge) and losses \cite{plasmons,leykam2016}, so long as the paraxial limit (or any other assumption responsible for topological protection) holds.  In Fig.~\ref{fig:stability}(b), we study the effects of a defect formed by detuning the depth of a single waveguide on the edge.  In the nontrivial lattice, the robustness of the solitons is found to be comparable to the linear edge states.  In the trivial lattice, the soliton survives for small defect strengths, but sufficiently strong defects cause it to abruptly destabilize (by leaving the ``stability band'' around $P_{\mathrm{es}}$).  The embedded solitons are thus inherently less robust than the linear topological edge states and the solitons of the nontrivial lattice.  On the other hand, both soliton types are robust against changes in the lattice shape; they are able to move around corners and missing sites without backscattering, and with much less dispersion compared to edge states of the linear PTI~\cite{supplementary}.

\begin{figure}

\includegraphics[width=\columnwidth]{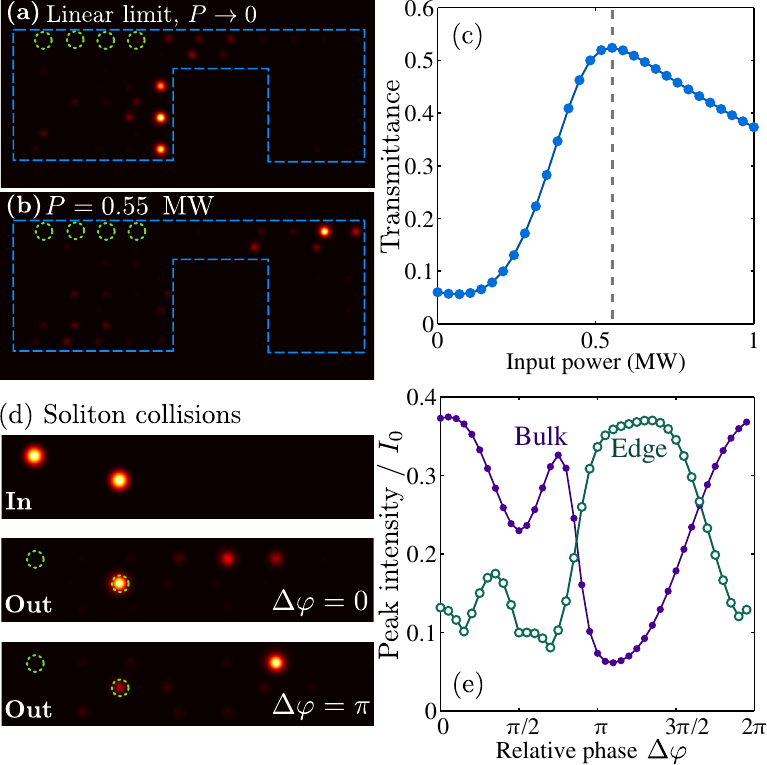}

\caption{(Color online) Nonlinear filtering and switching using edge solitons. (a)-(c) Power-dependent filtering by a narrow channel. The intensity profile at $z=8Z$ is plotted for (a) the linear limit $P \rightarrow 0$, and (b) the nonlinear regime at $P = 0.55$ MW.  The input light is injected uniformly on four waveguides marked by green circles, and the lattice boundary is marked by blue dashes.  (c) Transmittance through the channel (defined as the total intensity on the edge sites to the right of the channel) versus input power $P$.  (d)--(e) Optical switching by bulk-edge soliton collisions.  (d) Intensity profiles at $z = 0$ before the collision, and at $z = 5Z$ after the collision.  (e) Post-collision peak intensities in the bulk and edge waveguides, versus the relative phase $\Delta\varphi$ of the inputs.  }

\label{fig:switching_combined}

\end{figure}

We conclude with two examples of using solitons to perform nonlinear filtering and switching.  These will use the solitons of the nontrivial lattice, due to their greater stability.  In Fig.~\ref{fig:switching_combined}(a)--(c), we show how a narrow channel can act as a power-dependent filter.  The channel is one unit cell wide, connecting two wider strips.  In the linear regime [Fig.~\ref{fig:switching_combined}(a)], the edge states fail to pass through the channel, and are instead diverted, because topological protection is lost when there is a significant spatial overlap of modes on opposite edges.  Increasing the input power $P$ [Fig.~\ref{fig:switching_combined}(b)] causes the formation of a soliton that is sufficiently strongly localized to pass through the channel.  As shown in Fig.~\ref{fig:switching_combined}(c), the transmittance exhibits a $8.6$-fold variation as $P$ goes from 0 to 0.55 MW.

Next, Fig.~\ref{fig:switching_combined}(d)--(e) shows how a collision between a moving edge soliton and a stationary gap soliton can function as a nonlinear optical switch.  The stationary gap soliton is produced by exciting a waveguide one unit cell away from the edge \cite{supplementary}.  The two solitons are initialized with relative phase $\Delta\varphi$, and both have input power $P = P_c$ for which the soliton is maximally localized (see Fig.~\ref{fig:nontrivial_propagation}). We find that the result of the collision depends strongly on $\Delta\varphi$~\cite{soliton_collision,soliton_collision_2}.  Fig.~\ref{fig:switching_combined}(e) shows how the peak intensities on the edge and bulk sites, at $z = 5Z$ after the collision, vary with $\Delta\varphi$.  Certain choices of $\Delta\varphi$ allow us to almost completely destroy one of the solitons. 

In summary, we predict the existence of strongly-localized, mobile, and unidirectional edge solitons in experimentally feasible 2D photonic topological insulator lattices with Kerr nonlinearities.  Like topological edge states, the solitons move unidirectionally and can bypass corners and missing-site defects without backscattering.  The solitons in the nontrivial lattice inherit some of the linear edge states' robustness against perturbations such as waveguide detunings.  The trivial lattice supports topologically self-induced embedded solitons, which are unstable against perturbations and bending losses.  We have studied two simple examples of using solitons for nonlinear filtering and switching.  In future work, it would be interesting to explore using these solitons for nontrivial signal processing tasks; to determine whether the topological lattice design confers practical advantages over previously-studied solitonic lattices; and to look for similar nonlinear modes in other photonic, polaritonic, or phononic lattices.

\begin{acknowledgments}
We are grateful to T.~C.~H.~Liew, M.~C.~Rechtsman, C.~Soci, and B.~Zhang for helpful discussions. This research was supported by the Singapore National Research Foundation under grant No.~NRFF2012-02, and by the Singapore MOE Academic Research Fund Tier 2 grant MOE2015-T2-2-008.
\end{acknowledgments}

\clearpage

\begin{center}
  {\large \textbf{Supplemental Material}}
 
  {\large Edge Solitons in Nonlinear Photonic Topological Insulators}

  {\footnotesize Daniel Leykam and Y.~D.~Chong}
\end{center}

\makeatletter 
\renewcommand{\theequation}{S\arabic{equation}}
\makeatother
\setcounter{equation}{0}

\makeatletter 
\renewcommand{\thefigure}{S\@arabic\c@figure}
\makeatother
\setcounter{figure}{0}

\renewcommand{\theequation}{S\arabic{equation}}
\makeatother
\setcounter{equation}{0}

\makeatletter 
\renewcommand{\thefigure}{S\@arabic\c@figure}
\makeatother
\setcounter{figure}{0}

\noindent
This Supplemental Material is divided into three sections. First, we give a full description of the continuum model for the photonic lattice, as well as additional beam propagation simulation results showing the solitons moving around missing sites and corners as well as the behaviour of the system under defocusing nonlinearity.  Next, we introduce a nonlinear discrete (tight-binding) model that captures the key features of the system. Finally, we show that the discrete model supports self-consistent solutions consistent with the solitons found in the continuum model.

\section{Continuum model}

The lattice, shown in Fig.~1(a) of the main text, is described by the linear refractive index modulation~\cite{leykam2016}
\begin{multline}
n_L(x,y,z) = \Delta n_1 \sum_{nm} \Big\{ V_0(X_{n}^-, Y_{m}^-) \\
+ V_0 (X_{n+1/2}^+, Y^+_{m+1/2}) \Big\}, \label{eq:lattice_potential}
\end{multline}
where
\begin{align}
X_{n}^\pm(z)  &= x \pm x_0(z) - n d \\
Y_{m}^\pm(z) &=y \pm y_0(z) - m d.
\end{align}
Here, $\Delta n_1 = 2.7\times 10^{-3}$ is the modulation depth, $d=\sqrt{2}a$ is the lattice constant, and $a$ is the waveguide center-of-mass separation.  We assume the waveguide profiles $V_0$ are hypergaussian,
\be 
V_0(x,y) = \exp \left( - [(x^2+y^2)/\sigma^2]^3 \right),
\ee
with $\sigma = 4\,\mu$m.  The waveguide centers follow helices,
\be 
x_0(z) = R_0  \cos (\Omega z ), \quad y_0(z) =  R_0 \sin (\Omega z),
\ee
with radius $R_0 = 4\,\mu$m and pitch $Z = 2\pi/\Omega = 1$\,cm. These parameters are chosen to strike a balance between minimizing bending losses, allowing for strong nearest-neighbor couplings, and fitting many helix cycles within an experimentally feasible total array length ($\lesssim 10$~cm).

Except where otherwise noted, the beam propagation simulations use a lattice size of $N_x \times N_y = 12 \times 4$ unit cells, forming a quasi-one-dimensional strip. This strikes a balance between an array size that is physically feasible, while still having the ``bulk'' relatively well-defined. The full numerical grid size is $12 \times 6$ unit cells, with periodic boundary conditions along the $N_x$ axis. Absorbing boundary conditions are imposed along the $y$ axis edge, away from the edge waveguides.

The beam propagation simulations involve integrating the nonlinear Schr\"odinger equation using the symmetrized split-step method. Even without absorbing boundary conditions, the Hamiltonian is $z$-dependent and hence does not conserve ``energy'', so it cannot be used to estimate the numerical integration error.  To ensure the validity of the results, we simulate with various step sizes $dz$ until the results are no longer sensitive to $dz$.  For the propagation distances used in the main text ($z \simeq 10Z$), this requires $dz \approx Z/3000$. Note that the normalized ratio between the linear and nonlinear potentials is $n_2 I_0 / n_L \sim 0.1$ (with maximum peak intensities $|\psi|^2 \approx I_0$ occurring at $P_c$ in the trivial lattice), so we are in the regime of ``discrete'' solitons localized within the linear potential, rather than opposite regime of bulk solitons perturbed by a periodic potential, below the collapse threshold of the 2D nonlinear Schr\"odinger equation.

To find the local density of states in Fig.~2 of the main text, we calculate the Floquet spectrum of a semi-infinite strip of width $N_y=8$ unit cells ($N_x=1$ with twisted boundary conditions). Due to the finite size, the Bloch wave spectrum is discrete; we interpolate this finite level spacing using Lorentzians of width $1/N$~\cite{supp_LDOS}. The local density of states on the edge is obtained by integrating over the waveguide closest to the edge (one of the two sublattices).

Fig.~3 of the main text showed that propagation losses and single waveguide detuning defects can reduce the soliton power $P$; in particular, the embedded soliton abruptly breaks up when $P$ drops below  $P_{\mathrm{es}}$.  Here, we present further simulation results showing the effects of ``lattice shape deformations'', i.e.~corners and missing waveguides in the lattice.  These results show that the solitons are able to propagate around this class of defects without backscattering, similar to the topological edge states of the linear PTI.

\begin{figure}
\includegraphics[width=\columnwidth]{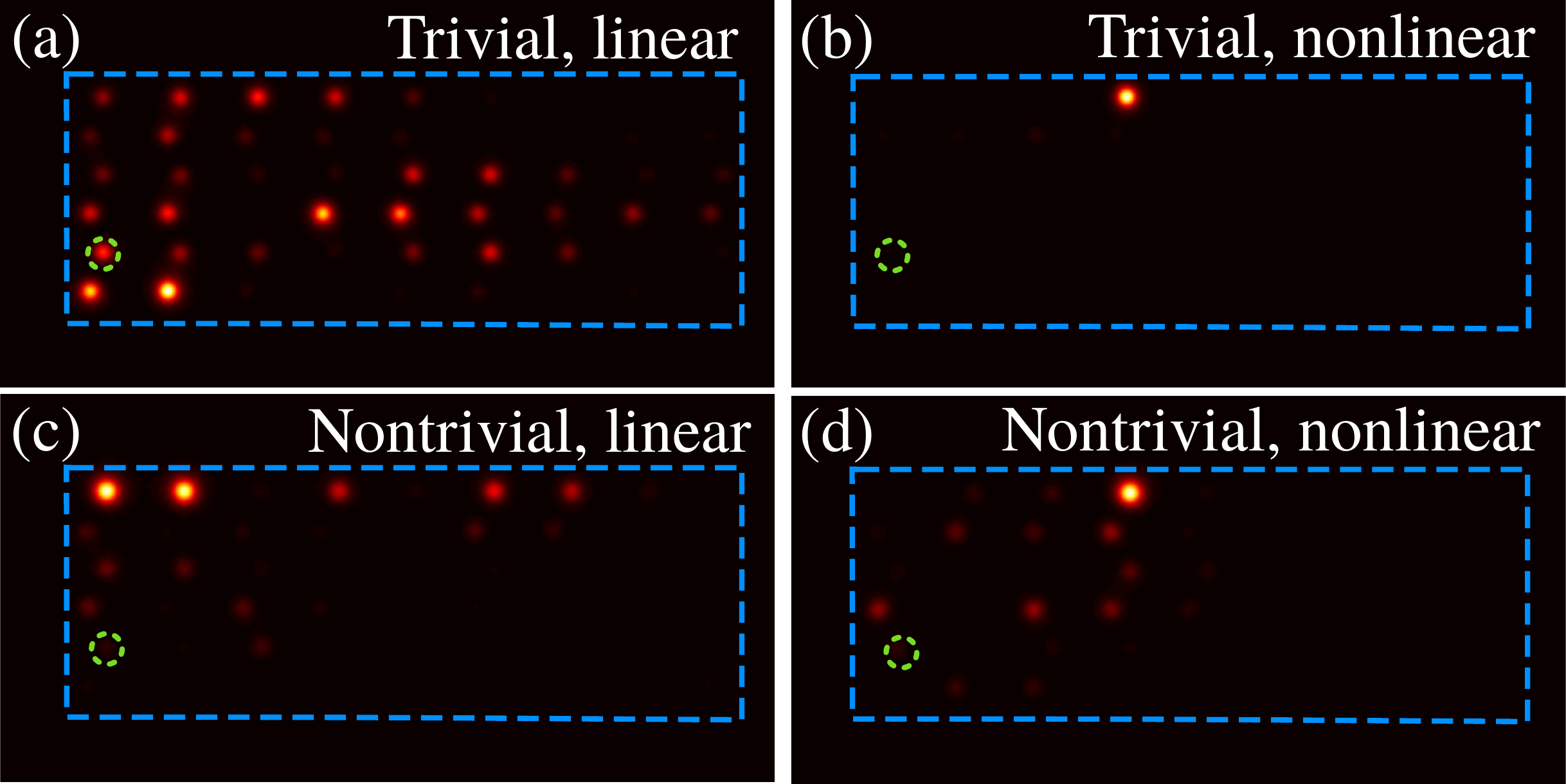}
\caption{Output intensity profiles after propagation $z=5Z$ around a corner, when waveguide circled in green is excited at the input. (a) Trivial linear output, no edge modes. (b) Trivial nonlinear output when the embedded soliton is excited, input power 1.6MW. (c) Nontrivial linear output. (d) Nontrivial nonlinear output, input power 0.5MW.}
\label{fig:corner}
\end{figure}

Fig.~\ref{fig:corner} shows propagation around a corner of a finite lattice ($N_x \times N_y = 10 \times 3$ cells).  Fig.~\ref{fig:corner}(a)--(b) shows the case where the lattice is topologically trivial in the limit limit.  There are no linear topological edge modes, so when the nonlinearity is negligible ($P \rightarrow 0$) an edge excitation simply diffracts into the bulk.  For $P = P_{\mathrm{es}}$, we produce an embedded soliton that is strongly localized to a single waveguide, and this soliton is able to travel around the corner without scattering into bulk modes.  Fig.~\ref{fig:corner}(c)--(d) shows the topologically nontrivial case.  For $P \rightarrow 0$, we see that the edge excitation couples into topological edge states that can propagate around the corner; however, there is significant dispersion, so that most of the wavepacket only moves 3 to 4 unit cells after propagation by $z=5Z$.  In the nonlinear limit, the soliton also propagates efficiently around the corner, but it moves by the full 5 unit cells.

\begin{figure}
\includegraphics[width=\columnwidth]{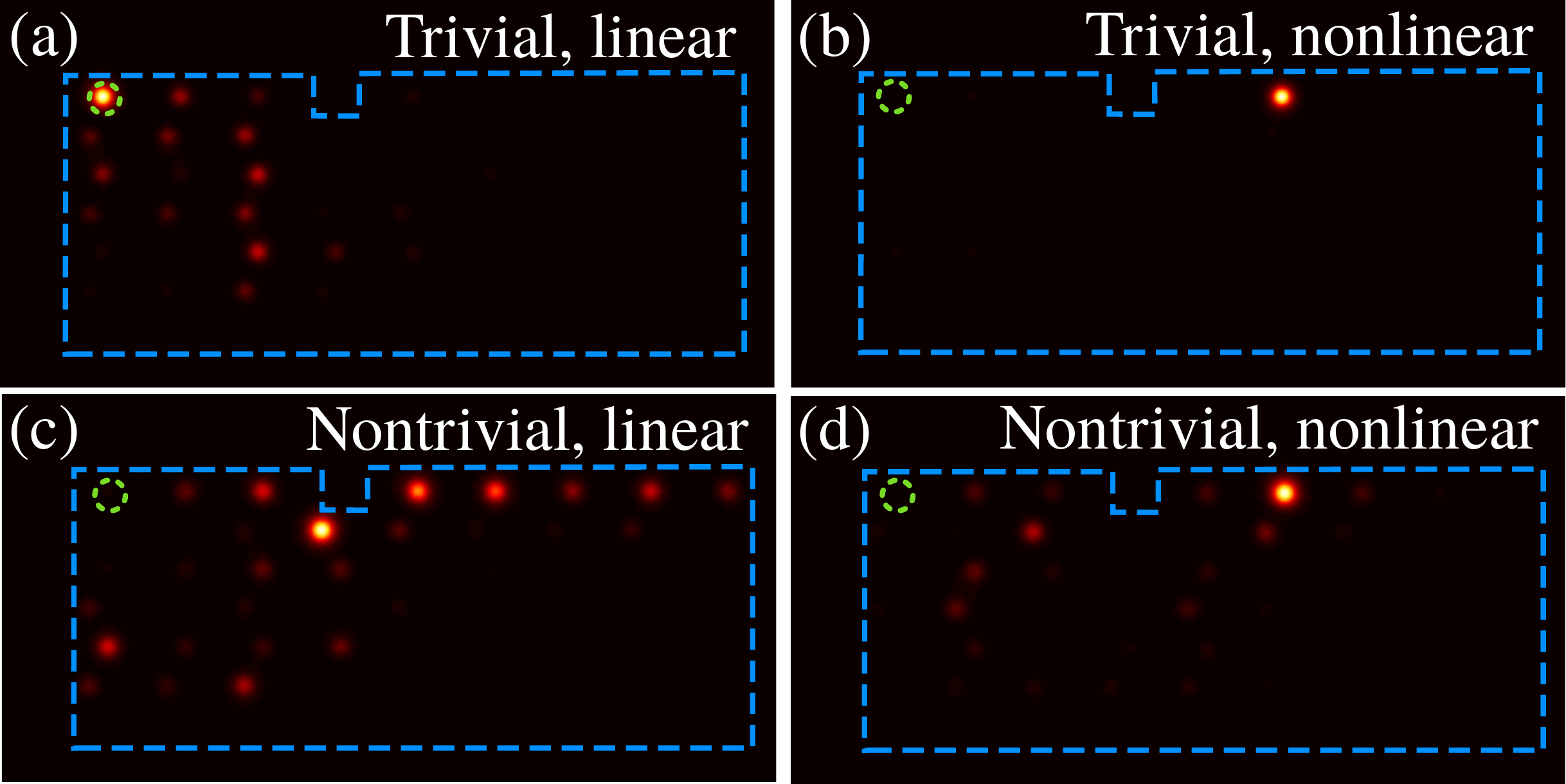}
\caption{Output intensity profiles after propagation $z=5Z$ around a missing edge waveguide (circled in blue), when waveguide circled in green is excited at the input. (a) Trivial linear output, no edge modes. (b) Trivial nonlinear output when the embedded soliton is excited, input power 1.6MW. (c) Nontrivial linear output. (d) Nontrivial nonlinear output, input power 0.5MW.}
\label{fig:defect}
\end{figure}

Fig.~\ref{fig:defect} shows the effects of removing one waveguide along the lattice edge.  In both the trivial and nontrivial phases, the strongly-localized solitons are able to travel around the defect, with minimal delay, and negligible scattering into the bulk or loss of localization.  By contrast, Fig.~\ref{fig:defect}(c) shows that the linear topological edge states experience significant delay and dispersion.  We thus conclude that the edge solitons are not only robust against lattice shape deformations, like the linear topological edge states, but experience much less distortion and less excess line delay.  However, we note that the defect-induced scattering into bulk modes in these examples is small but nonzero, and this does reduce the soliton lifetime compared to the defect-free case (Fig.~3(a) in the main text).

In the main text we considered a self-focusing nonlinearity, $n_2 > 0$. Photonic lattices with defocusing nonlinearities $n_2 < 0$ can also support localized solitons. In the tight binding approximation the $\beta \rightarrow -\beta$ symmetry of the linear spectrum enables a staggering transformation that maps soliton solutions for $+n_2$ to solutions for $-n_2$~\cite{signal_processing}. Based on this symmetry we expect the edge solitons to also appear in systems with defocusing nonlinearities. However, since the tight binding approximation does not exactly describe our system, this symmetry is not perfect and we do not expect the soliton properties to be exactly the same. Specifically, while both focusing and defocusing nonlinearities reduce the effective coupling strength by detuning the waveguides' propagation constants, a focusing nonlinearity increases the waveguide mode localization [increasing $\theta_{NL}$ in Eq. (3)], while a defocusing nonlinearity reduces it (decreasing $\theta_{NL}$). Hence for the defocusing nonlinearity we expect higher critical powers to observe the single site solitons.

\begin{figure}
\includegraphics[width=\columnwidth]{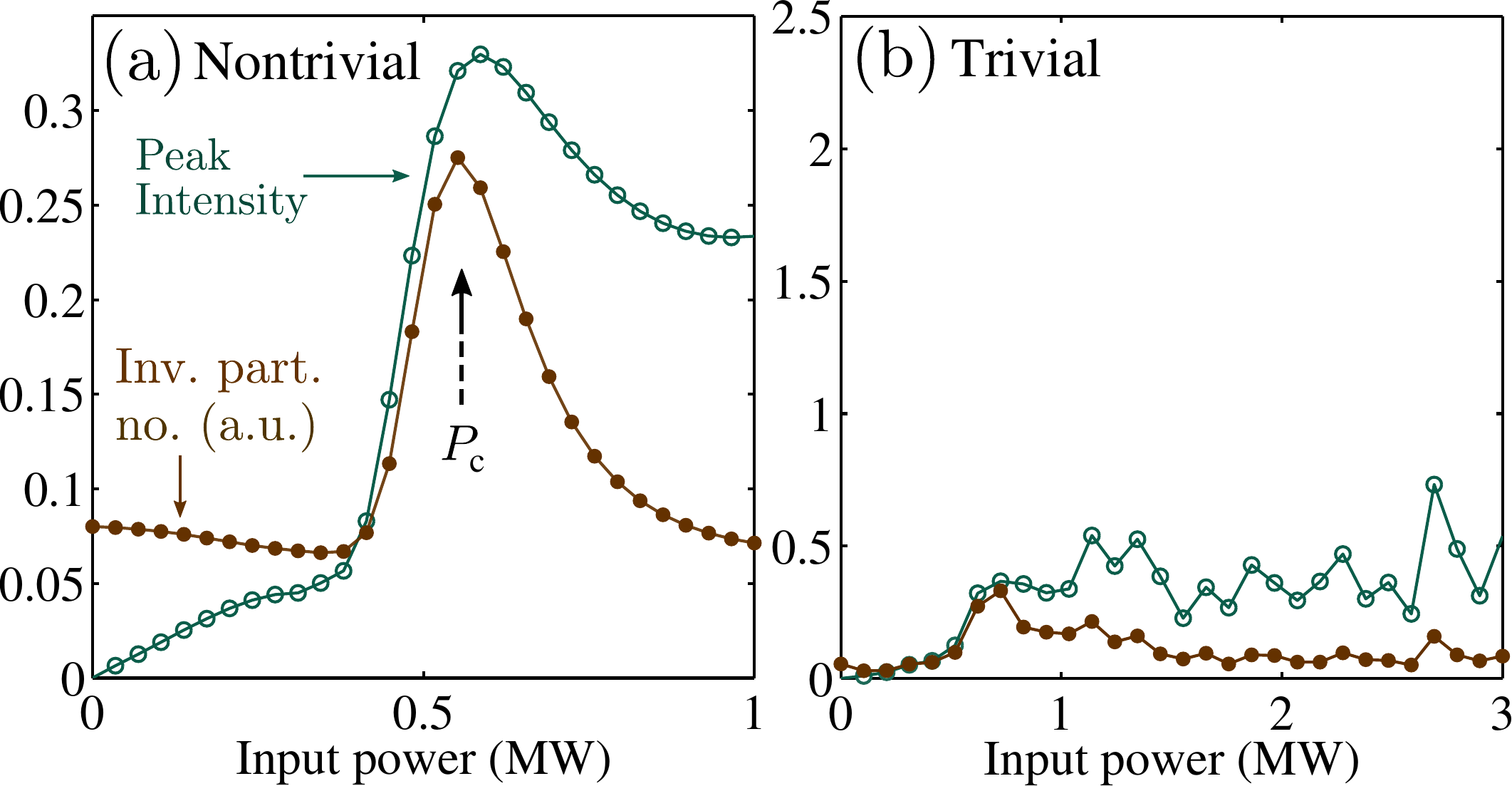}
\caption{Effect of defocusing nonlinearity on edge state formation. Peak intensity $I_{\mathrm{max}}$ and inverse participation number $\mathcal{P}^{-1}$ at $z = 5Z$ (i.e., after 5 helix cycles), versus input power $P$ in the nontrivial (a) and trivial (b) lattices, using the same parameters as in Figs.~1 and~2 of the main text respectively. (a) In the nontrivial lattice a strongly localized soliton forms at a higher threshold power compared to the focusing nonlinear case, $P_c \approx 0.6$MW. (b) In the trivial lattice the defocusing nonlinearity does not produce an embedded soliton.}
\label{fig:defocusing}
\end{figure}

To verify this reasoning we present in Fig.~\ref{fig:defocusing} the results of additional beam propagation simulations using the same parameters as in Figs.~1 and 2 of the main text, but reversing the sign of the nonlinear term $n_2$. In the nontrivial phase we observe the strongly localized soliton at a higher critical power $P_c \approx 0.6$MW where the output inverse participation number is maximized. Notice however that the peak intensity also displays a local maximum at $P_c$, in contrast to the focusing case Fig. 1(b), because the nonlinearity is now reducing the mode confinement. Therefore powers $\approx 0.5$MW are already sufficient to observe an asymmetry between focusing and defocusing nonlinearity. In the trivial lattice this asymmetry is even more pronounced: no embedded soliton is observed in the defocusing case, indicating that it becomes unstable above a critical power. To observe the embedded soliton in this case one should therefore adjust the linear lattice parameters to reduce threshold power $P_{\mathrm{ES}}$ to a level where the asymmetry between focusing and defocusing nonlinearities is less severe.

\section{Nonlinear discrete model}

We will now formulate a discrete nonlinear lattice model, and show that it can reproduce most of the soliton features found in the continuum model.  In the following, we normalize the lattice constant to $d=1$ and express the field as $\Psi = \psi(n,m)$, with integer values of $(n,m)$ corresponding to ``A'' sublattice sites and the ``B'' sublattice sites displaced by half a lattice constant. Each waveguide is coupled to its four neighbors in turn, with each coupling step described by scattering matrix $(\psi_A,\psi_B) = \hat{S} (\psi_A,\psi_B)$. The linear limit is completely characterized by coupling angle $\theta_0$, where
\be 
\hat{S}_0 = \left( \begin{array}{cc}  \cos \theta_0 & -i \sin \theta_0  \\ -i  \sin \theta_0 & \cos \theta_0 \end{array} \right). \label{eq:s0}
\ee
The evolution $\hat{U}$ over one modulation cycle is computed in four steps,
\begin{subequations}
\begin{align} 
\begin{pmatrix} \psi(n,m) \\ \psi(n+\frac{1}{2},m+\frac{1}{2}) \end{pmatrix} &= \hat{S}_0 \begin{pmatrix} \psi(n,m) \\ \psi(n+\frac{1}{2},m+\frac{1}{2}) \end{pmatrix} \\
\begin{pmatrix} \psi(n,m) \\ \psi(n+\frac{1}{2},m-\frac{1}{2}) \end{pmatrix} &= \hat{S}_0 \begin{pmatrix} \psi(n,m) \\ \psi(n+\frac{1}{2},m-\frac{1}{2}) \end{pmatrix} \\
\begin{pmatrix} \psi(n,m) \\ \psi(n-\frac{1}{2},m-\frac{1}{2}) \end{pmatrix} &= \hat{S}_0 \begin{pmatrix} \psi(n,m) \\ \psi(n-\frac{1}{2},m-\frac{1}{2}) \end{pmatrix} \\
\begin{pmatrix} \psi(n,m) \\ \psi(n-\frac{1}{2},m+\frac{1}{2}) \end{pmatrix} &= \hat{S}_0 \begin{pmatrix} \psi(n,m) \\ \psi(n-\frac{1}{2},m+\frac{1}{2}) \end{pmatrix}.
\end{align}
\end{subequations}
Although $\hat{U}$ is more simply expressed in Fourier space (see e.g.~Ref.~\cite{leykam2016}), this real-space description generalizes more easily to the nonlinear case.

In the linear limit the band edge modes at the Brillouin zone centre have quasienergies $\beta = \mathrm{Arg}( e^{\pm i 4 \theta_0} )$. The band gap $\Delta$ is given by the difference between these quasienergies, which cannot exceed $\pi$ because the quasienergies are only defined modulo $2\pi$. One can show that $\Delta = \frac{\pi}{2} - | 4 \mathrm{Mod}(\theta_0,\frac{\pi}{4} ) - \frac{\pi}{2} |$. Hence the coupling angle $\theta_0$ corresponding to the continuum model can be determined by choosing $\theta_0$ to reproduce the numerically-computed gap size. Trivial and nontrivial systems with the same gap size $\Delta$ are distinguished by the presence or absence of edge modes in a semi-infinite system.

Next, we replace $\hat{S}_0$ with a nonlinear scattering matrix $\hat{S}_{NL} (|\Psi|^2)$ describing a nonlinear two-port waveguide coupler.  For a bulk Kerr-type nonlinearity, there are two important effects. The first is a nonlinear detuning of the waveguide propagation constants $g |\psi|^2 \approx 1.4/$MW, where $g \propto n_2 / A_{\mathrm{eff}}$ is the effective Kerr coefficient and $A_{\mathrm{eff}} \sim (10\mu\mathrm{m})^2$ is the effective mode area~\cite{nonlinear_coupler}. A second and crucial effect is that the focusing (defocusing) nonlinearity makes the waveguides deeper (shallower), changing the mode confinement and therefore the evanescent coupling strength~\cite{nonlinear_coupler_2}. We model this as a power-dependent coupling strength, $C(I) = C_0 - C_{NL} I$, where $I = (|\psi_A|^2 + |\psi_B|^2)$ is the total power carried by the two coupled waveguides. Note that typically the nonlinearity cannot change the sign of the coupling, so this expression assumes sufficiently weak intensities $I < C_0 / C_{NL}$.  Taking both effects into account, the nonlinear scattering matrix can be written as
\begin{align}
  \begin{aligned}
\Psi_{\mathrm{out}} &= \exp \left[ -i \int_0^L  \left( \begin{array}{cc} g |\psi_A|^2 & C (I) \\ C (I) & g |\psi_B|^2  \end{array} \right) dz \right] \Psi_{\mathrm{in}} \\
&= \hat{S}_{\mathrm{NL}} \Psi_{\mathrm{in}},
  \end{aligned}
 \label{eq:nl_coupler}
\end{align}
where $L\approx 1.25$mm is the coupling length. This is an integrable model with exact solutions expressible in terms of elliptic functions~\cite{supp_coupler_solution}.  But since we are interested in the weakly nonlinear regime, below the self-trapping threshold, we instead use the split-step beam propagation method to integrate Eq.~\eqref{eq:nl_coupler}.

$L$ and $g$ can both be rescaled to 1 by an appropriate change of variables such that $C_0 \rightarrow \theta_0$ (estimated by fitting the linear spectrum) and the model behavior is entirely determined by the nonlinear coupling angle $\theta_{NL} = C_{NL}L$. We will assume that these rescaled units introduce the characteristic power scale $1$MW. Then $\theta_{NL}$ can be estimated from the embedded soliton critical power $P_c \approx 1.5$MW in Fig.~2 of the main text, by assuming that this corresponds to the power at which the effective coupling vanishes (i.e., the soliton becomes perfectly localized). We thus obtain
\begin{equation}
\theta_{NL} = (\theta_0 - \pi/2)/P_c \approx 0.25\pi / \text{MW},
\end{equation}
which is the value used in the following simulations (the precise value is not important, as long as it is not too small).

\section{Self-consistency of soliton solutions}

We can use a variant of the numerical method developed by Lumer \textit{et al.}~\cite{TI_solitons} to show that the Floquet lattice contains self-consistent soliton solutions.  We start with the simpler case of stationary gap solitons. We seek self-consistent stationary solutions of
\be 
\hat{U}(|\Psi|^2) \Psi = e^{-i \beta} \Psi,
\ee
where $\beta \in [-\pi,\pi]$ is the soliton's quasienergy. 

To find a solution, we fix the total power $P = \sum_{n,m} |\psi(n,m)|^2$, choose as a trial solution $\Psi_0$ a soliton perfectly localized to a single waveguide, and apply the following iteration procedure: to obtain the updated profile $\Psi_{j+1}$, we first calculate $\hat{U}(|\Psi_j|^2)$ using the beam propagation method described above.  While performing the nonlinear propagation, we simultaneously carry out linear propagation of a basis of states in the nonlinearly-induced potential; this yields the matrix elements of $\hat{U}(|\Psi_j|^2)$ in this basis. Diagonalizing this matrix gives the eigenmodes of the induced potential, normalized to have total power $P$. The eigenmode with largest overlap with $\Psi_j$ is then used as the updated profile $\Psi_{j+1}$. When the two are identical, the profile forms a soliton, a localized eigenmode trapped by its self-induced potential. We iterate until the error $\epsilon$ is sufficiently small, $|\Psi_{j}-\Psi_{j-1}|^2/P < \epsilon$. The propagation constant $\beta$ is given by the corresponding eigenvalue of $\hat{U}(|\Psi_j|^2)$

\begin{figure}

\includegraphics[width=\columnwidth]{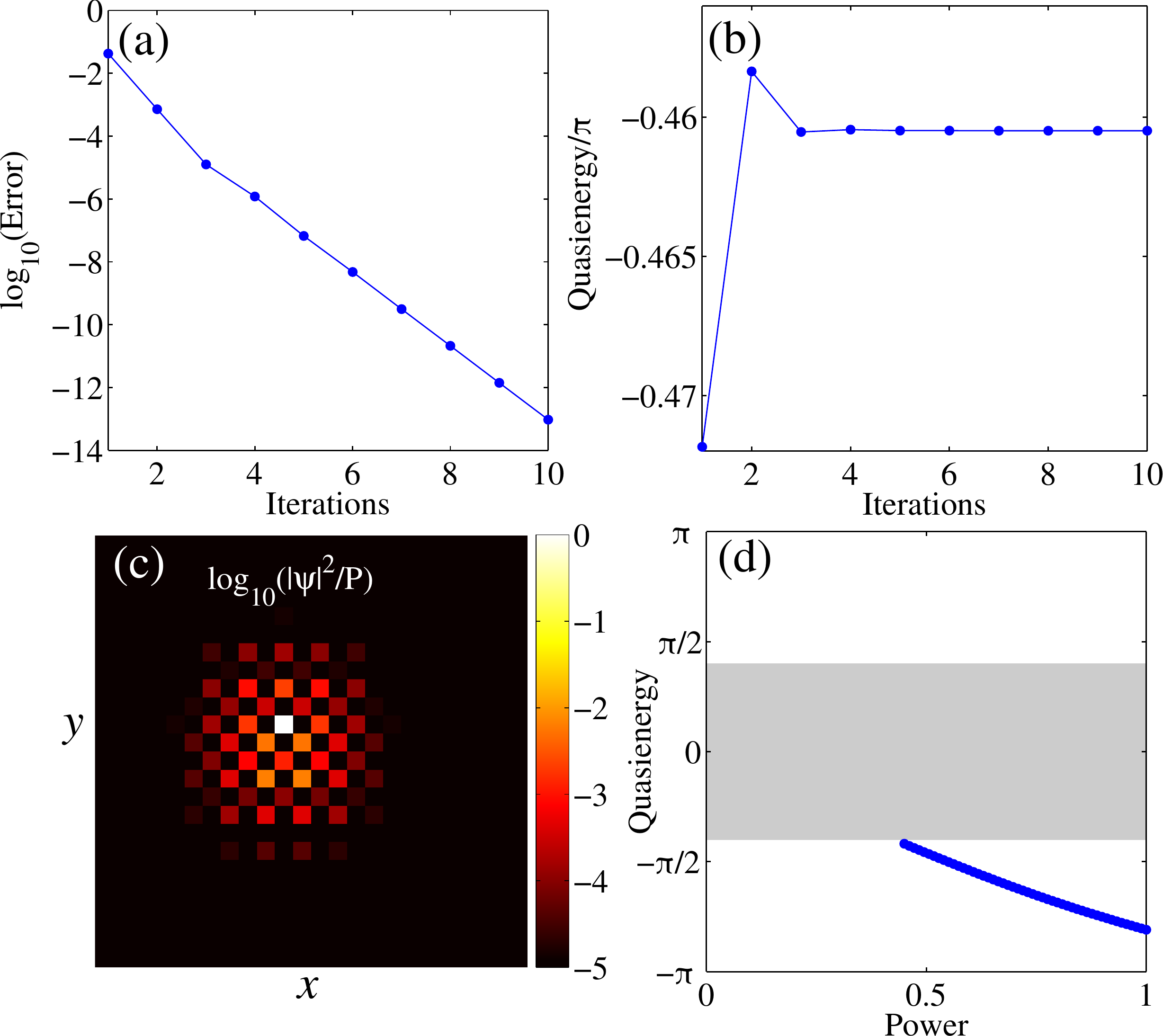}

\caption{Bulk soliton in the anomalous Floquet insulator phase. (a)--(b) Convergence of the iteration procedure to a soliton with power $P=0.5$. (c) Corresponding soliton intensity profile, log scale, illustrating exponential localization. (d) Power diagram showing the bulk soliton family, bifurcating from linear Bloch wave spectrum (shaded gray region) at nonzero threshold power $P_{min}\approx 0.45$.}

\label{fig:bulk_soliton_example}

\end{figure}

Fig.~\ref{fig:bulk_soliton_example} illustrates the application of this procedure to obtain stationary bulk solitons in the nontrivial lattice. We use a lattice size of $N=12\times12$ unit cells, and set an error tolerance of $\epsilon = 10^{-10}$. Convergence is achieved within about 10 iterations. The soliton's intensity profile displays the expected exponential localization. Performing beam propagation on the obtained soliton, we verify its stable propagation exceeding 50 modulation cycles, even when slightly perturbed. Furthermore, during each modulation cycle the soliton circulates in a loop, resembling the behavior observed in Ref.~\cite{TI_solitons}. This demonstrates that the bulk solitons involve a topologically self-induced ``inner edge''.  (In the trivial phase, the bulk solitons do \emph{not} circulate, but instead remain centred at a single waveguide throughout the modulation cycle.)  Fig.~\ref{fig:bulk_soliton_example}(d) shows the power diagram for this family of gap solitons, which bifurcates from the linear Bloch wave spectrum with a nonzero threshold power $P_{\mathrm{min}} \approx 0.45$, as expected for solitons in 2D (we also obtain a family of gap solitons in the trivial phase).  These numerical results verify the applicability of the Floquet self-consistency method to the staggered-helix lattice.

To obtain the traveling solitons, we combine the evolution $\hat{U}(|\Psi|^2)$ with a translation $\hat{T}$.  In other words, we seek self-consistent solutions to
\be 
\hat{T}\hat{U}(|\Psi|^2) \Psi = e^{-i \beta} \Psi, \label{eq:travelling}
\ee
where $\hat{T} \psi(n,m) = \psi(n+1,m)$ corresponds to a soliton traveling one unit cell per modulation cycle. Note that while the spectrum of $\hat{U}(|\Psi|^2)$ is typically banded, supporting gap solitons, the spectrum of $\hat{T} \hat{U}$ is not, reflecting the fact that there are always resonant traveling linear waves in the bulk; cf.~Fig.~2(d) in the main text. Therefore the self-consistent solutions to Eq.~\eqref{eq:travelling} always involve a localized soliton with power $P$ on top of a nonzero background of delocalized linear waves. When there are linear edge modes, this background is strongly suppressed, but it still exists, giving the edge solitons a finite (but large) lifetime even in the absence of bending losses.

\begin{figure}

\includegraphics[width=\columnwidth]{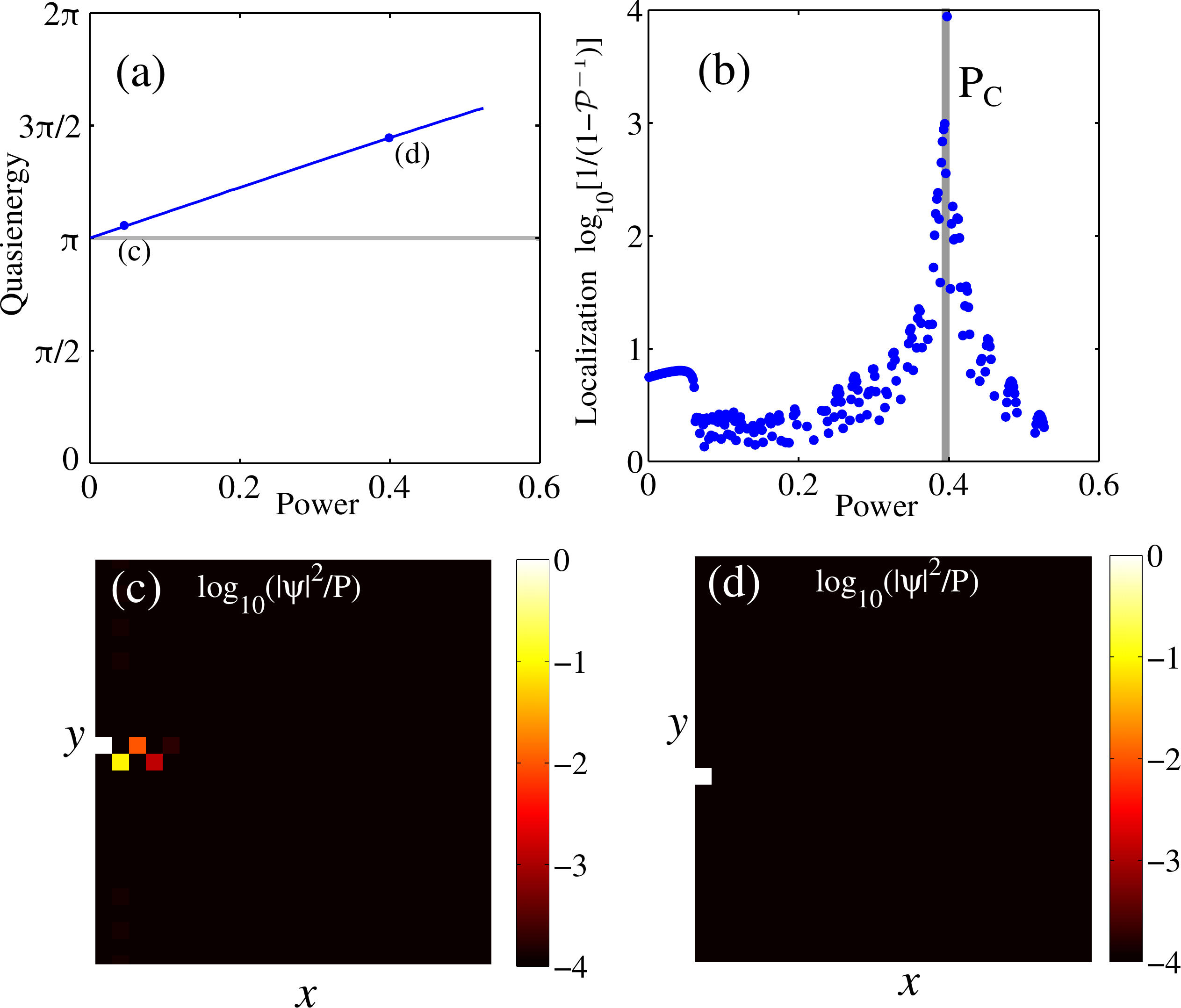}

\caption{Moving edge solitons in the anomalous Floquet insulator lattice. (a) Converged soliton quasienergy versus soliton power, showing that the solitons form a family bifurcating from the zero-power limit. The gray line shows the (flat) spectrum of the linear topological edge states. Filled circles correspond to the intensity profiles in (c)--(d). (b) Soliton localization versus power, as measured by the deviation of the inverse participation number $\mathcal{P}^{-1}$ from unity. Near-perfect localization occurs at critical power $P_c \approx 0.4$. (c) Intensity profile of a low power soliton, localized slightly more strongly to edge than a linear edge mode. (d) Intensity profile of a soliton at $P_c$, showing perfect localization on the edge.}

\label{fig:nontrivial_soliton_example}

\end{figure}

Fig.~\ref{fig:nontrivial_soliton_example} shows self-consistent edge soliton solutions occurring in the nontrivial lattice.  Like the stationary bulk solitons, these solitons form a continuous family; however, since they are quasi-1D solitons (localized to the edge), they bifurcate from the linear limit, unlike the bulk solitons which bifurcate at a nonzero threshold power. Note that the discrete model does not account for the nonzero dispersion of the edge states, so the linear traveling wave spectrum in Fig.~\ref{fig:nontrivial_soliton_example}(a) is perfectly flat and all the solitons are perfectly localized along the edge. On the other hand, the degree of localization into the bulk is power-dependent, as shown in Fig.~\ref{fig:nontrivial_soliton_example}(b)-(d). To measure the soliton localization, we use the inverse participation number $\mathcal{P}^{-1} = \sum_{n,m}|\Psi(n,m)|^4 / (\sum_{n,m} |\Psi(n,m)|^2)^2$. When $\Psi$ is perfectly localized to a single site $\mathcal{P} = 1$; otherwise $\mathcal{P} < 1$. Fig.~\ref{fig:nontrivial_soliton_example}(b) shows that near-perfect localization occurs at the critical power $P_c\approx 0.4$ corresponding to vanishing effective coupling rate. 

\begin{figure}

\includegraphics[width=\columnwidth]{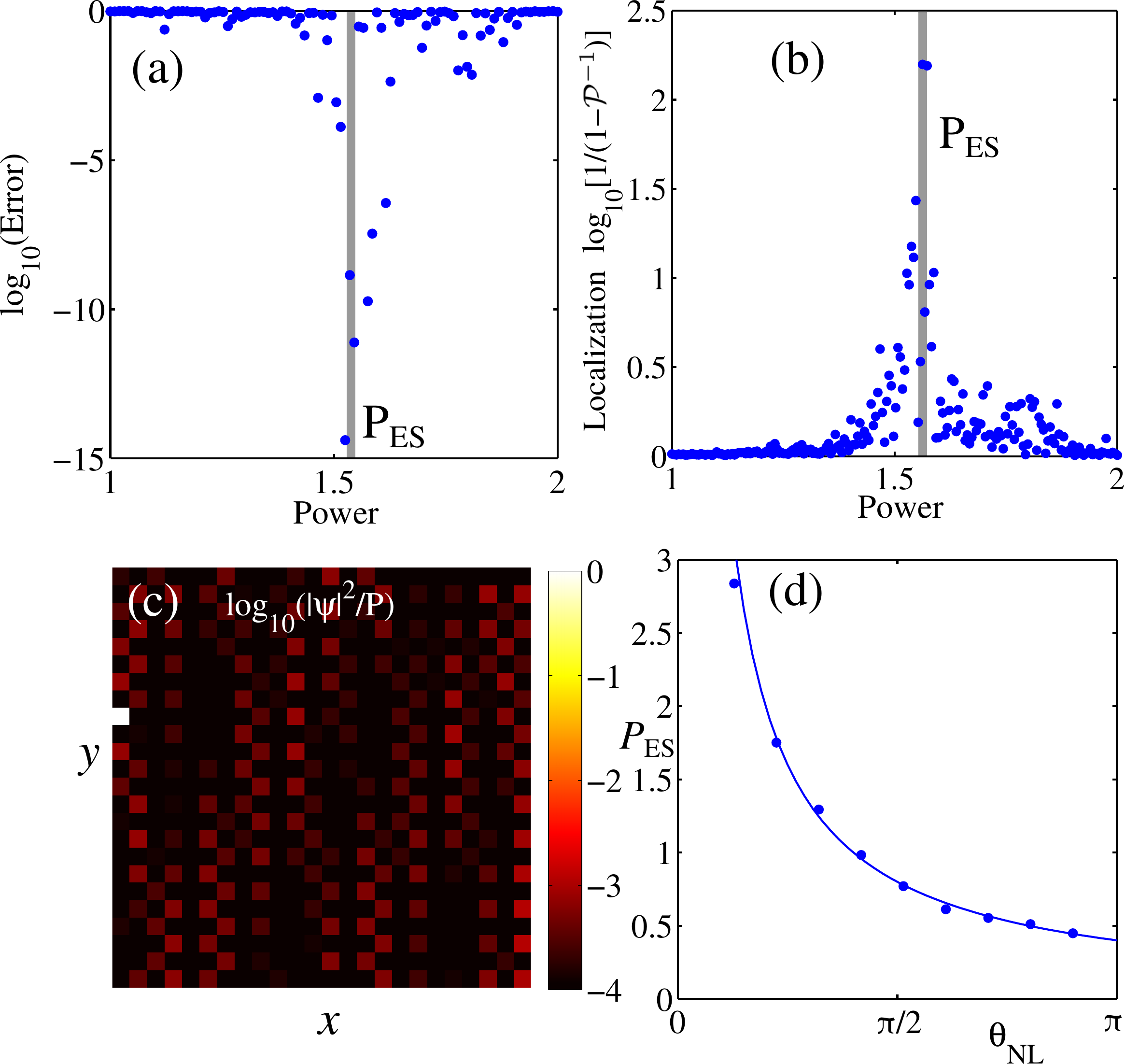}

\caption{Embedded traveling soliton in the trivial phase. (a) Error after 20 iterations versus soliton power. Convergence to stationary solution only occurs at the critical power $P_{\mathrm{ES}}$ (gray line). (b) Localization of $\Psi$ measured by the deviation of the inverse participation number $\mathcal{P}^{-1}$ from unity. The delocalized linear background only decouples from the edge state at $P_{\mathrm{ES}}$. (c) Intensity profile of the solution as $P_{\mathrm{ES}}$. Note the weak, non-decaying background present due to the finite accuracy of the numerical solution. (d) $P_{\mathrm{ES}}$ as a function of the nonlinear coupling angle $\theta_{NL}$. Points: numerical solution. Curve: the estimate $P_{\mathrm{ES}} = (\theta_0 - \pi/2)/\theta_{NL}$.}

\label{fig:trivial_soliton_example}

\end{figure}

Fig.~\ref{fig:trivial_soliton_example} shows the behavior of the trivial lattice, which is very different. For most choices of the power $P$, the iteration procedure no longer converges to a stationary solution. Convergence only occurs at a critical power $P_{\mathrm{es}}$, at which the soliton becomes almost perfectly localized to a single waveguide (the finite background in Fig.~\ref{fig:trivial_soliton_example}(c) is due to the finite accuracy of the numerical solution). This discrete soliton spectrum is characteristic of embedded solitons.  Fig.~\ref{fig:trivial_soliton_example}(d) shows the dependence of $P_{\mathrm{es}}$ on the nonlinear coupling angle $\theta_{NL}$, which becomes very large as $\theta_{NL} \rightarrow 0$.

\begin{figure}

\includegraphics[width=0.8\columnwidth]{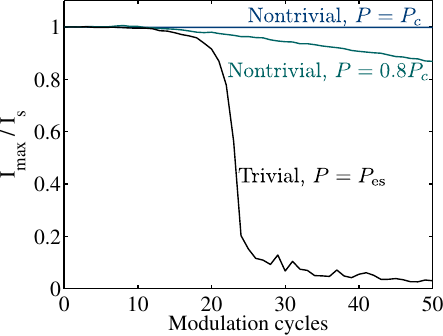}

\caption{Dynamical stability of the numerically-obtained solitons when weakly perturbed, characterized by the peak intensity $I_{\mathrm{max}}$ relative to the soliton profile peak intensity $I_s$. }

\label{fig:discrete_soliton_stability}

\end{figure}

We examine in Fig.~\ref{fig:discrete_soliton_stability} the dynamical stability of the obtained solitons by weakly perturbing the soliton profiles and performing beam propagation simulations for $z=50$ modulation cycles. We perturb the solitons with a random background field of peak intensity $0.001I_s$, where $I_s$ is the soliton's peak intensity. If the soliton is stable it will be practically unaffected by this weak perturbation and remain an almost stationary solution, such that the peak intensity $I_{\mathrm{max}}(z) = \mathrm{max}|\psi(n,m, z)|^2 \approx I_s$. If the soliton is unstable the perturbation will grow, destroying the soliton. 

The soliton in the trivial phase is unstable, as expected for an embedded soliton, and disintegrates after tens of modulation cycles under the weak perturbation (the exact lifetime dependent on the seeding noise). On the other hand, the nontrivial soliton is stable near $P_c$, remaining practically unchanged after 50 modulation cycles. Further away from $P_c$, perturbations to the nontrivial solitons lead to a slow radiation of energy into the bulk; in this sense, the solitons are actually quasi-stable. Nevertheless, they are practically unchanged for experimentally-feasible propagation distances of tens of modulation cycles.

One might wonder if the self-consistency method can also be applied to the full continuum model, by projecting the continuous field profile onto a finite basis formed by the waveguide modes, and using the truncated basis method of Ref.~\cite{leykam2016}. There are two problems that make this difficult in practice. Firstly, the potential induced by the soliton breaks the translational symmetry of the lattice. Thus one must run the eigensolver on a full 2D lattice, rather than a single unit cell (Bloch wave spectrum) or 1D strip (edge spectrum), significantly increasing the computation time for each iteration step. Secondly, the truncated basis itself becomes dependent on the power and soliton profile, such that the state generated after propagation by one modulation cycle is not necessarily a superposition of the original basis; hence, truncation to a small basis may no longer be accurate, unless one performs an additional relaxation within each iteration step to determine the ``best'' truncated basis that captures the dynamics. These limitations make the self-consistency scheme impractical in the continuum model.

\end{document}